\documentclass[11pt,a4paper]{article}
\usepackage[T1]{fontenc}
\usepackage[utf8]{inputenc}
\usepackage{lmodern,microtype}
\usepackage[margin=27mm]{geometry}
\usepackage{amsmath,amssymb,amsthm,mathtools,booktabs}
\usepackage{tikz,pgfplots}
\pgfplotsset{compat=1.18}
\usepackage[colorlinks=true,linkcolor=blue!45!black,citecolor=blue!45!black,urlcolor=blue!45!black]{hyperref}
\numberwithin{equation}{section}
\newtheorem{theorem}{Theorem}[section]
\newtheorem{lemma}[theorem]{Lemma}
\newtheorem{proposition}[theorem]{Proposition}

\theoremstyle{remark}
\newtheorem{remark}[theorem]{Remark}
\newcommand{\dd}{\,\mathrm d}
\newcommand{\kin}{C_{\mathrm{kin}}}
\newcommand{\ks}{k_*}
\newcommand{\Xs}{\mathcal X_*}
\newcommand{\Ps}{\mathcal P}
\newcommand{\ak}{\mathfrak a_k}
\newcommand{\bk}{\mathfrak b_k}
\newcommand{\Wz}{W_0}
\title{Kinetic moment estimates and a one-sided Lyapunov barrier\protect\\
for the relativistic fermionic Michie--King equation}
\author{%
Dorota Bors\textsuperscript{1}\quad and\quad Robert Sta\'nczy\textsuperscript{2}\\[0.6em]
\small\textsuperscript{1}Department of Differential Equations and Computer Science,\\
\small Faculty of Mathematics and Computer Science,\\
\small University of \L{}\'od\'z, \L{}\'od\'z, Poland\\[0.35em]
\small\textsuperscript{2}Mathematical Institute, University of Wroc\l{}aw, Wroc\l{}aw, Poland}
\date{Research manuscript --- 4 October 2026}
\begin{document}
\maketitle
\begin{abstract}
We study the cutoff Ruffini--Arg\"uelles--Rueda (RAR), or relativistic
fermionic Michie--King, equation of state along an equilibrium family
with fixed Tolman--Klein invariants. Three kinetic moments give exact
formulas for both the pressure-to-density ratio and the derivative of
pressure. Cauchy--Schwarz yields a global relation between these two
quantities. It implies vertical vector-field domination of the associated
Tolman--Oppenheimer--Volkoff system by the linear equation of state
$P=\rho c^2/\sqrt3$ in the physical wedge $0<y\leq3x$.
We construct a filled, one-sided version of the Bors--Sta\'nczy
Lyapunov barrier and check every boundary. This controls all outward
excursions, including those after a lower exit from a Lyapunov sublevel.
For regular configurations the resulting explicit Lambert bound is
$2Gm(r)/(rc^2)<0.6879212521$. Using the diagonal where this ratio has a
stationary value sharpens the bound to $0.6734385929$.
The argument is conditional on a regular TOV configuration and does not
require a separate finite-radius existence theorem. We also identify and
correct a logarithmic coefficient in the previously used explicit
linear-comparator formula.
\end{abstract}

\medskip
{\small
\par}

\section*{Introduction}
\addcontentsline{toc}{section}{Introduction}

How much mass can be enclosed within a given radius in a static sphere?
The answer depends on how strongly the matter's pressure responds to a
change in density. In general relativity, the balance between pressure
and gravity is described by the Tolman--Oppenheimer--Volkoff equations.
We study a model of fermions with an energy cutoff: particles with enough
energy to escape do not contribute to the density or pressure of the
equilibrium.

The pressure in this model is not simply proportional to the density.
An estimate based on the distribution of particle energies lets us
compare the equilibrium equations with those for a simpler, linear
pressure law. We follow the enclosed mass and local density as the radius
increases. The comparison helps construct a boundary that the resulting
curve cannot cross. This is the Lyapunov barrier used in the proof.

The quantity we bound is \emph{compactness}: at a given radius,
it is the Schwarzschild radius associated with the enclosed mass divided
by that radius. The bound holds throughout the regular equilibrium
and is approximately \(0.67344\). If the density drops to zero at a
finite surface, the surface radius must exceed a fixed multiple of the
Schwarzschild radius of the total mass. That multiple is about \(1.485\).
Theorem~\ref{thm:main} states the exact bound and its assumptions.

\section{Scope, conventions, and principal result}

Throughout, $\rho$ denotes energy density divided by $c^2$, rather than
particle-number density or rest-mass density. Thus $\rho c^2$ is the
energy density, and $P$ is the physical pressure. We write
\begin{equation}\label{eq:wa}
 w(\rho)=\frac{P(\rho)}{\rho c^2},\qquad
 a(\rho)=\frac{P'(\rho)}{c^2}.
\end{equation}
The derivative in \eqref{eq:wa} is along the one-parameter equilibrium
curve specified below. It is not asserted to be an adiabatic derivative
under independently varying thermodynamic parameters.

The linear TOV phase system and its Lyapunov construction are developed
in \cite{BS-TOV,SB-cloud}; the kinetic cutoff model is described in
\cite{RAR}. Our calculation starts from the parametrization in slides
33--34 of \cite{ETAMM}. All estimates and the nonlinear barrier transfer
needed here are proved explicitly.

\begin{theorem}[Compactness bound]\label{thm:main}
Fix $\kappa>0$ and $0\leq\chi<\infty$, and let $P(\rho)$ be the cutoff
kinetic equation of state of Section~\ref{sec:kinetic}. Suppose that
$(m,\rho)$ is a regular, static, isotropic TOV configuration, with
$m(0)=0$ and $0<\rho(0)<\infty$. On its connected positive-density
interval, assume initially that $2Gm(r)/(rc^2)<1$.
Then, throughout that interval,
\begin{equation}\label{eq:mainbound}
 \frac{2Gm(r)}{rc^2}<\Xs,
 \qquad \Xs=0.6734385928381108\ldots.
\end{equation}
Here $\Xs$ is defined exactly as the unique root in $(z_*,1)$ of
\begin{equation}\label{eq:mainroot}
 U_*(\Xs)+J_*(\Xs)=E_*,
\end{equation}
where the functions and constants are given in
Section~\ref{sec:numbers}. A simpler explicit bound is
\begin{equation}\label{eq:lambertmain}
 \frac{2Gm(r)}{rc^2}<X_*^{\mathrm L},\qquad
 X_*^{\mathrm L}=0.6879212520282978\ldots.
\end{equation}
If the configuration has a finite surface $R$ with $\rho(R)=0$, then
the same inequalities hold at the surface, with $M=m(R)$. In particular,
\begin{equation}\label{eq:radius}
 R>\frac{2}{\Xs}\frac{GM}{c^2}
   =2.9698327676340697\ldots\frac{GM}{c^2}.
\end{equation}
The inequalities preclude continuation to a horizon $2Gm/(rc^2)=1$
from the regular positive-density branch.
\end{theorem}

The exact equations, rather than rounded decimal values, define the
constants. The theorem concerns every radius, so it also applies to a
core or to an arbitrarily truncated positive-density configuration.
For a finite surface the proof below actually gives strict inequality
there. No unimodality of $m(r)/r$ is assumed.

\section{The cutoff kinetic family}\label{sec:kinetic}

Let $m_f$ be the particle mass and $h_{\mathrm P}$ Planck's constant.
Introduce the dimensionless kinetic energy and momentum
\begin{equation}
 q=\frac{\varepsilon}{m_fc^2},\qquad
 \sigma(q)=\sqrt{q(q+2)}=\frac{p}{m_fc}.
\end{equation}
Let $u$ denote the dimensionless cutoff potential, so that the symbols
$x,y$ remain available for the TOV phase variables. For
$0<u<\kappa^{-1}$ set
\begin{equation}\label{eq:distribution}
 \beta(u)=\frac{\kappa}{1-\kappa u},\qquad
 q_c(u)=\beta(u)u,\qquad
 F(z)=\frac{1-e^z}{1+\chi e^z}\quad(z\leq0),
\end{equation}
and define
\begin{equation}
 f_u(q)=F\!\left(\frac{q}{\beta(u)}-u\right)
 \quad(0\leq q\leq q_c(u)),\qquad f_u(q)=0\quad(q>q_c(u)).
\end{equation}
For the physical fermionic family $\chi>0$; allowing $\chi=0$ includes a
classical limiting cutoff distribution. The equilibrium invariants are
$\beta^{-1}+u=\kappa^{-1}$ and $\chi=e^{u-\theta}$, where $\theta$ is
the chemical degeneracy parameter. Thus both $\kappa$ and $\chi$ remain
fixed while $u$ varies along a configuration.

With spin degeneracy two, put
\begin{equation}
 \kin=\frac{8\pi m_f^4c^3}{h_{\mathrm P}^3}.
\end{equation}
The density and normalized pressure $\Pi=P/c^2$ are
\begin{align}
 \rho(u)&=\kin\int_0^{q_c(u)}f_u(q)(q+1)^2\sigma(q)\dd q,
       \label{eq:rhoenergy}\\
 \Pi(u)&=\frac{\kin}{3}\int_0^{q_c(u)}f_u(q)\sigma(q)^3\dd q.
       \label{eq:penergy}
\end{align}
These formulas follow directly from
$\rho=(2m_f/h_{\mathrm P}^3)\int f(1+q)\dd^3p$ and
$P=(2/(3h_{\mathrm P}^3))\int f p v\dd^3p$, since
$v=c\sigma/(q+1)$ and
$p^2\dd p=m_f^3c^3\sigma(q)(q+1)\dd q$.

For comparison with slide~34, use $q=\beta\xi$ in
\eqref{eq:rhoenergy}--\eqref{eq:penergy}:
\begin{align}
 \rho(u)&=\sqrt2\kin\beta^{3/2}
 \int_0^u \xi^{1/2}(1+\beta\xi)^2
       (1+\beta\xi/2)^{1/2}F(\xi-u)\dd\xi,\label{eq:sliderho}\\
 P(u)&=\eta\beta^{5/2}
 \int_0^u \xi^{3/2}(2+\beta\xi)^{3/2}F(\xi-u)\dd\xi,
 \qquad \eta=\frac{8\pi m_f^4c^5}{3h_{\mathrm P}^3}.
 \label{eq:slidep}
\end{align}
The pressure normalization is fixed here by the kinetic formula, so
that no unspecified constant enters the derivative estimates.
Appendix~\ref{app:slide34} collects the complete formulas and both
endpoint asymptotics of slide~34 in a standalone form.
Appendix~\ref{app:hydrodynamic} derives the same slope bound directly
from the density and pressure integrals.

\section{Exact moment identities and the global slope estimate}

Define three positive moments, each at the same value of $u$:
\begin{equation}\label{eq:moments}
 A=\int_0^{q_c}f_u\sigma^3\dd q,\qquad
 B=\int_0^{q_c}f_u\sigma\dd q,\qquad
 C=\int_0^{q_c}\frac{f_u}{\sigma}\dd q.
\end{equation}
The last integral is finite because $\sigma(q)\sim\sqrt{2q}$ at zero.
Using $(q+1)^2=\sigma^2+1$ gives
\begin{equation}\label{eq:momentdensity}
 \rho=\kin(A+B),\qquad \Pi=\frac{\kin}{3}A,
 \qquad w=\frac{A}{3(A+B)}.
\end{equation}

\begin{lemma}[Differentiation identities]\label{lem:derivatives}
For $0<u<\kappa^{-1}$,
\begin{align}
 \rho_u&=\kin\beta(4A+5B+C),\label{eq:rhou}\\
 \Pi_u&=\kin\beta\left(\frac43A+B\right),\label{eq:pu}\\
 a&=\frac{\frac43A+B}{4A+5B+C}.\label{eq:exacta}
\end{align}
In particular, $\rho_u>0$, so the family defines a single-valued
equation of state wherever its density is positive.
\end{lemma}
\begin{proof}
At fixed $q$, the argument $z=q/\beta-u$ satisfies
$z_u=-(q+1)$ and $z_q=\beta^{-1}$. Hence
\begin{equation}\label{eq:transport}
 \partial_u f_u=-\beta(q+1)\partial_q f_u.
\end{equation}
The moving-endpoint terms vanish because $f_u(q_c)=F(0)=0$.
Integration by parts yields
\[
 \frac{\rho_u}{\kin}
 =\beta\int_0^{q_c}f_u\frac{\mathrm d}{\mathrm dq}
       \big((q+1)^3\sigma\big)\dd q,
 \qquad
 \frac{\Pi_u}{\kin}
 =\frac{\beta}{3}\int_0^{q_c}f_u\frac{\mathrm d}{\mathrm dq}
       \big((q+1)\sigma^3\big)\dd q.
\]
The lower boundary terms also vanish. Since
$\sigma'=(q+1)/\sigma$, the two differentiated kernels are
$4\sigma^3+5\sigma+\sigma^{-1}$ and
$4\sigma^3+3\sigma$, respectively. Their integrable singularity at
zero justifies the integration by parts by first integrating from a
positive lower endpoint and then taking its limit. Division proves
\eqref{eq:exacta}.
\end{proof}

\begin{proposition}[Kinetic inequalities]\label{prop:kinetic}
For every positive density in this family,
\begin{equation}\label{eq:bounds}
 0<w<\frac13,\qquad 0<a<\frac13,
 \qquad
 a<\frac{3w(1+w)}{1+9w}.
\end{equation}
\end{proposition}
\begin{proof}
The first bound follows from $B>0$ in \eqref{eq:momentdensity}.
For the second, observe that
$3(\frac43A+B)=4A+3B<4A+5B+C$.
Apply Cauchy--Schwarz in the measure $f_u(q)\dd q$:
\begin{equation}\label{eq:cs}
 B^2=\left(\int\sigma^{3/2}\sigma^{-1/2}f_u\dd q\right)^2
 \leq AC.
\end{equation}
Equality would require $\sigma^2$ to be constant almost everywhere in
this measure. As $f_u>0$ on an interval, equality is impossible.
Substituting $C>B^2/A$ into \eqref{eq:exacta} gives
\[
 a<\frac{\frac43A+B}{4A+5B+B^2/A}.
\]
Writing $r=A/B$ and $w=r/[3(r+1)]$ simplifies the right-hand side to
$3w(1+w)/(1+9w)$.
\end{proof}

\begin{remark}
The non-strict version of the last inequality depends only on the moment
representation and transport identity. The particular form of $F$ is
needed for the endpoint asymptotics, and its positive interval support
gives strictness.
\end{remark}

\section{Endpoint asymptotics and enthalpy}\label{sec:asymptotics}

As $u\downarrow0$, $\beta\to\kappa$ and
$F(\xi-u)=(u-\xi)/(1+\chi)+O(u^2)$ uniformly for $0\leq\xi\leq u$.
The identities
\[
 \int_0^u\xi^{1/2}(u-\xi)\dd\xi=\frac4{15}u^{5/2},\qquad
 \int_0^u\xi^{3/2}(u-\xi)\dd\xi=\frac4{35}u^{7/2}
\]
therefore imply
\begin{align}
 \rho(u)&\sim\rho_0u^{5/2},&
 \rho_0&=\frac{4\sqrt2\kin\kappa^{3/2}}{15(1+\chi)},\label{eq:smallrho}\\
 P(u)&\sim P_0u^{7/2},&
 P_0&=\frac{8\sqrt2\kin c^2\kappa^{5/2}}{105(1+\chi)}.
 \label{eq:smallp}
\end{align}
Consequently
\begin{equation}\label{eq:polytrope}
 P(\rho)\sim P_0(\rho/\rho_0)^{7/5},\qquad
 w\sim\frac{2\kappa}{7}u,\qquad
 a\sim\frac{2\kappa}{5}u\sim\frac75w.
\end{equation}
The slope asymptotic follows by applying the same rescaling to
\eqref{eq:rhou}--\eqref{eq:pu}, rather than by differentiating an
unqualified asymptotic equivalence.

For the other endpoint set $\delta=1-\kappa u\downarrow0$ and
$v=\delta q$. Then $q_c=(1-\delta)/\delta$ and
\[
 f_u(v/\delta)=F((v-1+\delta)/\kappa).
\]
Dominated convergence gives
\begin{equation}\label{eq:highmoments}
 A\sim I_3\delta^{-4},\qquad
 B\sim I_1\delta^{-2},\qquad
 I_j=\int_0^1v^jF((v-1)/\kappa)\dd v>0.
\end{equation}
Also $0\leq f_u\leq1$, and hence
\[
 C\leq\int_0^{q_c}\frac{\dd q}{\sqrt{q(q+2)}}
   =\operatorname{arcosh}(1+q_c)=O(\log(1/\delta)).
\]
It follows that $\rho\to\infty$, $w\to1/3$, $a\to1/3$ and
\begin{equation}
 P(\rho)\sim\frac{c^2}{3}\rho.
\end{equation}
Together with $\rho_u>0$ this shows that the EOS is defined for all
$\rho>0$.

There is an exact enthalpy identity. Define dimensionless enthalpy by
\begin{equation}
 H(\rho)=\int_0^\rho\frac{P'(t)}{tc^2+P(t)}\dd t.
\end{equation}
Since \eqref{eq:pu} and \eqref{eq:momentdensity} give
$\Pi_u=\beta(\rho+\Pi)$,
\begin{equation}\label{eq:enthalpy}
 H(u)=-\log(1-\kappa u),\qquad H_u=\beta(u).
\end{equation}
In particular, $H$ tends to zero at vacuum and is finite at every finite
central density.

\section{Why the linear comparator \texorpdfstring{$k=1/3$}{k=1/3} does not follow}

The sufficient comparison condition previously proposed for $k=1/3$
was
\begin{equation}\label{eq:failed}
 8a\leq(1+w)(1+3w).
\end{equation}
It is not a global inequality for the present EOS. Indeed, put
$r=A/B$ and $\gamma=C/B$. The exact identities give
\[
 w=\frac{r}{3(r+1)},\qquad
 a=\frac{4r+3}{3(4r+5+\gamma)}.
\]
Substitution into \eqref{eq:failed} shows that it is equivalent to
\begin{equation}
 \gamma\geq\frac{2r+3}{2r+1}.
\end{equation}
But \eqref{eq:highmoments} implies $r\to\infty$ and $\gamma\to0$,
whereas the displayed lower bound tends to one.

More precisely, with $\epsilon=B/A\to0$ and $C/A=o(\epsilon)$,
\begin{align}
 w&=\frac13-\frac{\epsilon}{3}+O(\epsilon^2),\\
 a&=\frac13-\frac{\epsilon}{6}+o(\epsilon),\\
 \frac{(1+w)(1+3w)}8
   &=\frac13-\frac{\epsilon}{4}+O(\epsilon^2).
\end{align}
Thus the left slope exceeds the proposed upper slope by
$\epsilon/12+o(\epsilon)>0$ near the relativistic endpoint.
This disproves that sufficient EOS inequality; it does not by itself
disprove every possible comparison argument using $k=1/3$.

\section{The TOV system in mass--density variables}\label{sec:tov}

The physical TOV equations are
\begin{equation}\label{eq:tov}
 \frac{\mathrm dm}{\mathrm dr}=4\pi r^2\rho,
 \qquad
 \frac{\mathrm dP}{\mathrm dr}
 =-\frac{G(\rho+P/c^2)(m+4\pi r^3P/c^2)}
         {r^2(1-2Gm/(rc^2))}.
\end{equation}
Choose an arbitrary reference length $r_0>0$ and let
\begin{equation}\label{eq:milne}
 s=\log(r/r_0),\qquad
 x=\frac{2Gm(r)}{rc^2},\qquad
 y=\frac{8\pi Gr^2\rho(r)}{c^2}.
\end{equation}
Primes below mean $\mathrm d/\mathrm ds$. Direct substitution yields
\begin{equation}\label{eq:nonlinear}
 x'=y-x,\qquad
 y'=F_{\mathrm R}(x,y;\rho)
 =y\left[2-\frac{1+w}{2a}\frac{x+wy}{1-x}\right].
\end{equation}
The EOS coefficients depend on
$\rho=c^2y/(8\pi Gr_0^2e^{2s})$, so the planar system is generally
nonautonomous. Its enthalpy equation is
\begin{equation}\label{eq:hevolution}
 H'=-\frac{x+wy}{2(1-x)},\qquad
 u'=-\frac{x+wy}{2\beta(1-x)}.
\end{equation}

For $0<x<1$ and positive density, TOV and $P'>0$ imply
$\rho_r<0$. Therefore
\begin{equation}\label{eq:wedge}
 m(r)=4\pi\int_0^r t^2\rho(t)\dd t
       >\frac{4\pi}{3}r^3\rho(r),\qquad
 0<y<3x.
\end{equation}
At a regular center with density $\rho_c>0$,
\begin{equation}\label{eq:center}
 x\sim\frac{8\pi G\rho_c}{3c^2}r^2,\qquad
 y\sim\frac{8\pi G\rho_c}{c^2}r^2,
 \qquad y/x\longrightarrow3.
\end{equation}
Thus the orbit starts at the origin as $s\to-\infty$ and immediately
lies in the physical wedge. The integral estimate
\eqref{eq:wedge} is regular at $r=0$ when interpreted by this limit.

\section{The optimal comparator furnished by the moment bound}

For the linear EOS $P=k\rho c^2$, $k>0$, define
\begin{equation}\label{eq:linear}
 F_k(x,y)=y\left[2-\frac{1+k}{2k}\frac{x+ky}{1-x}\right].
\end{equation}
Assume $w\leq k$ and set $t=y/x\in(0,3]$. The inequality
$F_{\mathrm R}\leq F_k$ is equivalent to
\begin{equation}
 a\leq\frac{k(1+w)}{1+k}\frac{1+wt}{1+kt}.
\end{equation}
The fraction decreases with $t$, so it suffices that
\begin{equation}\label{eq:sufficient}
 a\leq\frac{k(1+w)(1+3w)}{(1+k)(1+3k)}.
\end{equation}

\begin{proposition}[Uniform field domination]\label{prop:comparison}
For $\ks=1/\sqrt3$,
\begin{equation}\label{eq:domination}
 F_{\mathrm R}(x,y;\rho)<F_{\ks}(x,y)
 \quad\text{if }0<x<1,\quad0<y\leq3x,\quad\rho>0.
\end{equation}
\end{proposition}
\begin{proof}
By Proposition~\ref{prop:kinetic}, \eqref{eq:sufficient} follows if
\begin{equation}\label{eq:optimization}
 \frac{3w}{(1+9w)(1+3w)}
 \leq\frac{k}{(1+k)(1+3k)}.
\end{equation}
For $0\leq w\leq1/3$, the left side has maximum
$1-\sqrt3/2$, attained at $w=1/(3\sqrt3)$; its derivative has the
sign of $1-27w^2$. For $k>0$, the right side has the same maximum,
attained at $k=1/\sqrt3$; its derivative has the sign of $1-3k^2$.
Moreover $w<1/3<\ks$, as required. The strict moment inequality
in \eqref{eq:bounds} makes the resulting field inequality strict.
\end{proof}

\begin{remark}[Meaning of optimality]
$\ks$ is the unique constant selected by demanding
\eqref{eq:optimization} for the entire interval $0\leq w\leq1/3$
using only the Cauchy--Schwarz moment estimate. This is optimality of
that sufficient envelope, not a claim that the actual cutoff EOS has
no sharper comparator. Distribution-specific moment estimates may
improve it.
\end{remark}

\section{The normalized linear Lyapunov function}\label{sec:lyapunov}

For any $k>0$, write
\begin{equation}
 \ak(x)=\frac{4k-(1+5k)x}{2k(1-x)},\qquad
 \bk(x)=\frac{1+k}{2(1-x)},\qquad
 F_k=\ak(x)y-\bk(x)y^2.
\end{equation}
The positive equilibrium is $(z,z)$, where
\begin{equation}\label{eq:z}
 z=\frac{4k}{1+6k+k^2}.
\end{equation}
Introduce
\begin{equation}\label{eq:HB}
 h_k=\frac{1+5k}{2k},\qquad
 b_k=h_k(1-z)=\frac{(1+5k)(1+k)^2}{2k(1+6k+k^2)},
\end{equation}
and the nonnegative potentials
\begin{align}
 U_k(x)&=h_k(z-x)+b_k\log\frac{1-z}{1-x},\quad0\leq x<1,
          \label{eq:U}\\
 J_k(y)&=y-z-z\log(y/z),\quad y>0.\label{eq:J}
\end{align}
Then
\begin{equation}\label{eq:gradients}
 U_k'(x)=\frac{h_k(x-z)}{1-x}=z\bk(x)-\ak(x),\qquad
 J_k'(y)=1-\frac zy.
\end{equation}
Both potentials vanish at $z$ and have their unique minimum there.
For $V_k=U_k+J_k$, direct differentiation along the linear system gives
\begin{equation}\label{eq:lyapid}
 (1-x)V_k'
 =-h_k(x-z)^2-\frac{1+k}{2}(y-z)^2.
\end{equation}
The conventional Lyapunov function in slide~32 is $2V_k$, up to its
normalizing constant. Its logarithmic coefficient satisfies
$z\delta_k=b_k$, with
$8k^2\delta_k=(5k+1)(1+k)^2$.

Along \eqref{eq:nonlinear},
\begin{equation}\label{eq:transfer}
 V_k'=-\frac{h_k(x-z)^2+(1+k)(y-z)^2/2}{1-x}
       +\left(1-\frac zy\right)(F_{\mathrm R}-F_k).
\end{equation}
Under field domination this derivative is nonpositive for $y\geq z$,
and strictly negative there except possibly at $(z,z)$. For $y<z$ the
comparison term has the opposite sign. We therefore use only the upper
Lyapunov arc and fill the region below it.

\section{A filled barrier and the complete boundary check}\label{sec:barrier}

The comparator nullcline intersects the tangent $y=3x$ at
\begin{equation}\label{eq:xi}
 \xi_k=\frac{4k}{1+8k+3k^2},\qquad Y_k=3\xi_k.
\end{equation}
We have $0<\xi_k<z<Y_k$. Let
\begin{equation}\label{eq:E}
 E_k=J_k(Y_k)=Y_k-z-z\log(Y_k/z)>0,
\end{equation}
and let $X_k^{\mathrm L}\in(z,1)$ be the unique root of
\begin{equation}\label{eq:Xroot}
 U_k(X_k^{\mathrm L})=E_k.
\end{equation}
It exists because $U_k$ increases strictly from zero to infinity on
$(z,1)$. Define the continuously differentiable function
\begin{equation}\label{eq:truncated}
 \Ps_k(x,y)=U_k(\max\{x,z\})+J_k(\max\{y,z\}),\qquad y\geq0.
\end{equation}
Despite the maxima in this formula, $\Ps_k$ is $C^1$: the one-sided
derivatives match because $U_k'(z)=J_k'(z)=0$. In particular,
\eqref{eq:truncated} is finite at $y=0$.
Our filled barrier is
\begin{equation}\label{eq:filled}
 \mathcal B_k=\{0\leq x<1,\ y\geq0:\ \Ps_k(x,y)\leq E_k\},
\end{equation}
considered inside the physical wedge $y\leq3x$.

For $x\leq z$, its upper boundary is $y=Y_k$.
For $z\leq x\leq X_k^{\mathrm L}$, the upper boundary is
$y=\gamma_k(x)\geq z$, implicitly defined by
\begin{equation}
 U_k(x)+J_k(\gamma_k(x))=E_k.
\end{equation}
The upper arc ends at $(X_k^{\mathrm L},z)$; a vertical boundary then
continues down to $(X_k^{\mathrm L},0)$. The lower half of a full
Lyapunov oval is not a boundary of this filled region.

\begin{lemma}[One-sided barrier]\label{lem:barrier}
Suppose a positive $C^1$ orbit satisfies $x'=y-x$,
$y'=F_{\mathrm R}\leq F_k$ in $0<x<1$, $0<y\leq3x$, and approaches
$(0,0)$ at its regular center. If the domination is strict, it remains
in the strict sublevel $\Ps_k<E_k$ for every finite interior radius.
In particular $x<X_k^{\mathrm L}$.
\end{lemma}
\begin{proof}
Near the center $x,y<z$, and $\Ps_k=0<E_k$. We verify the derivative
on every possible positive-density boundary point with
$\Ps_k=E_k$.

\emph{Horizontal cap.} If $x\leq z$, the boundary is $y=Y_k>z$.
The wedge implies $x\geq Y_k/3=\xi_k$. Also
\[
 F_k(x,Y_k)=\frac{Y_k}{2k(1-x)}
       \{4k-(1+5k)x-k(1+k)Y_k\}\leq0,
\]
because the numerator vanishes at $x=\xi_k$ and decreases with $x$.
Thus $\Ps_k'=(1-z/Y_k)F_{\mathrm R}<0$ under strict domination.
This proves the cap, including its junction at $x=z$.

\emph{Upper Lyapunov arc.} If $x\geq z$ and $y\geq z$, then
$\Ps_k=V_k$. Formula \eqref{eq:transfer} gives $\Ps_k'<0$ on its
level $E_k>0$, since $(z,z)$ is not on that level. This includes
the endpoint $(X_k^{\mathrm L},z)$.

\emph{Right vertical side.} If $y\leq z$, the boundary equation is
$U_k(x)=E_k$, hence $x=X_k^{\mathrm L}>z$. There
\[
 \Ps_k'=U_k'(X_k^{\mathrm L})(y-X_k^{\mathrm L})<0.
\]
The orbit points left even though the original $V_k$ may increase.

These are all points of the positive-density level
$\Ps_k=E_k$. At a first contact with that level, the derivative of
$\Ps_k$ would be nonnegative, contradicting the strict inequalities
just proved. Hence no contact occurs. Since $J_k(\max\{y,z\})\geq0$
and $U_k$ is increasing for $x\geq z$, we have $x<X_k^{\mathrm L}$.
\end{proof}

For completeness, the remaining physical or internal curves require
the following checks.
\begin{enumerate}
\item On $y=3x$, with $\Phi=y-3x$, the nonlinear field itself gives
\begin{equation}
 \Phi'= -\frac{3(1+w)(1+3w)x^2}{2a(1-x)}<0.
\end{equation}
Thus the physical wedge has the correct orientation independently of
the comparator. Density monotonicity already gives its strict form.
\item At $x=0$, formally $x'=y\geq0$. In the physical wedge, $x=0$
forces $y=0$; the regular-center expansion specifies entry into the
positive branch. Mass positivity gives $x>0$ at all interior radii.
\item The boundary $y=0$ is a vacuum endpoint for the present variables.
No division by $a(0)=0$ is used there. A finite surface is treated by
the interior limit. In the exterior vacuum, $m=M$, $y=0$, and
$x'=-x<0$.
\item The boundary $x=1$ is excluded by
$x<X_k^{\mathrm L}<1$ on the maximal initially horizon-free interval.
Continuity prevents a first horizon from being reached.
\item Neither the diagonal $y=x$ nor the comparator nullcline is a
required invariant boundary of $\mathcal B_k$. An orbit may cross
either one in the admissible region. Repeated excursions are included
in the same first-contact proof.
\end{enumerate}

\begin{remark}[Why a lower exit needs an additional argument]
For the full Lyapunov function $V_k$, the safe derivative is only known
for $y\geq z$. If $x\geq z$ and $y<z$, then $x'<0$, so an exit in
this part cannot increase $x$. That observation alone does not control
a later return. The filled function \eqref{eq:truncated}, together with
the horizontal cap, supplies the missing global return control. The
orbit can leave the ordinary Lyapunov sublevel while remaining inside
$\mathcal B_k$.
\end{remark}

\begin{proposition}[Diagonal refinement]\label{prop:diagonal}
Under the strict hypotheses of Lemma~\ref{lem:barrier}, let
$\mathcal X_k\in(z,X_k^{\mathrm L})$ be the unique solution of
\begin{equation}\label{eq:diagroot}
 U_k(\mathcal X_k)+J_k(\mathcal X_k)=E_k.
\end{equation}
Then $x<\mathcal X_k$ at every interior radius.
\end{proposition}
\begin{proof}
The left side of \eqref{eq:diagroot} is zero at $z$ and strictly
increases for $x>z$; at $X_k^{\mathrm L}$ it exceeds $E_k$.
If $x$ first reached $\mathcal X_k$ from below, then $x'\geq0$,
so $y\geq x=\mathcal X_k>z$. Since $J_k$ is increasing for $y\geq z$,
\[
 \Ps_k(x,y)=U_k(x)+J_k(y)
 \geq U_k(\mathcal X_k)+J_k(\mathcal X_k)=E_k,
\]
contrary to the strict sublevel established in the lemma.
\end{proof}

\section{Explicit constants and the corrected mass--radius estimate}
\label{sec:numbers}

Apply the barrier with $k=\ks=1/\sqrt3$. In this section an asterisk
denotes evaluation at $\ks$. The equilibrium and cap parameters are
\begin{align}
 z_*&=\frac{18-4\sqrt3}{23}=0.4813824682488909\ldots,\\
 \xi_*&=\frac{8-2\sqrt3}{13}=0.3489152603740189\ldots,\\
 Y_*&=3\xi_*=1.0467457811220566\ldots.
\end{align}
Moreover
\begin{equation}
 h_* =\frac{5+\sqrt3}{2},\qquad
 b_*=h_*(1-z_*),\qquad E_*=Y_*-z_*-z_*\log(Y_*/z_*).
\end{equation}
To invert $U_k(X)=E_k$, put
\begin{equation}
 D_k=h_kz+b_k\log(1-z),\qquad
 \alpha_k=\frac{h_k}{b_k}=\frac1{1-z}.
\end{equation}
Then $U_k(x)=-h_kx-b_k\log(1-x)+D_k$. With $t=1-X$, its root satisfies
\[
 -\alpha_kt\exp(-\alpha_kt)
 =-\alpha_k\exp\left(-\alpha_k-\frac{E_k-D_k}{b_k}\right).
\]
Therefore the right-hand root is
\begin{equation}\label{eq:Lambert}
 X_k^{\mathrm L}
 =1+\frac1{\alpha_k}\Wz\!\left(
  -\alpha_k\exp\left[-\alpha_k-\frac{E_k-D_k}{b_k}\right]
 \right).
\end{equation}
The principal branch $\Wz$ is required because the root has $X>z$,
equivalently $0<\alpha_k(1-X)<1$. For $E_k>0$ the argument belongs
to $(-e^{-1},0)$.

The constants used in the calculation are
\begin{center}
\begin{tabular}{@{}lr@{}}
\toprule
Quantity & Value at $k=1/\sqrt3$\\
\midrule
$h_*$ & $3.3660254037844386$\\
$b_*$ & $1.7456797867222159$\\
$D_*$ & $0.4741521689299105$\\
$E_*$ & $0.1914353915526654$\\
$\alpha_*$ & $1.9282032302755092$\\
$X_*^{\mathrm L}$ & $0.6879212520282978$\\
$\Xs$ & $0.6734385928381108$\\
$2/X_*^{\mathrm L}$ & $2.9073095126849337$\\
$2/\Xs$ & $2.9698327676340697$\\
\bottomrule
\end{tabular}
\end{center}
For the sharper root, the scalar equation can also be written as
\begin{equation}\label{eq:scalarsharp}
 (h_k-1)(z-X)
 +b_k\log\frac{1-z}{1-X}-z\log\frac{X}{z}=E_k.
\end{equation}
Its strict monotonicity for $X>z$ makes bisection unambiguous.
The displayed roots were computed with 50-decimal arithmetic by
bracketing \eqref{eq:Xroot} and \eqref{eq:scalarsharp}; the Lambert
representation is an analytic equivalent, not a reliance on an
unspecified branch of a numerical library.

\begin{proof}[Proof of Theorem~\ref{thm:main}]
Equations \eqref{eq:center} and \eqref{eq:wedge} give the regular
initial entry and the physical wedge. Proposition~\ref{prop:comparison}
provides strict domination. Lemma~\ref{lem:barrier} gives
\eqref{eq:lambertmain}, and Proposition~\ref{prop:diagonal} gives
\eqref{eq:mainbound}. Since this applies on the entire initially
horizon-free positive-density interval, a first contact with $x=1$
cannot occur.

At a finite surface, $x$ is continuous and $y\to0$. Because $x(R)>0$,
we have $x'=y-x<0$ in a sufficiently short interval before the surface.
Thus $x(R)$ is strictly below a preceding interior value, which is
already strictly below $\Xs$. Rearranging the surface compactness
inequality proves \eqref{eq:radius}.
\end{proof}

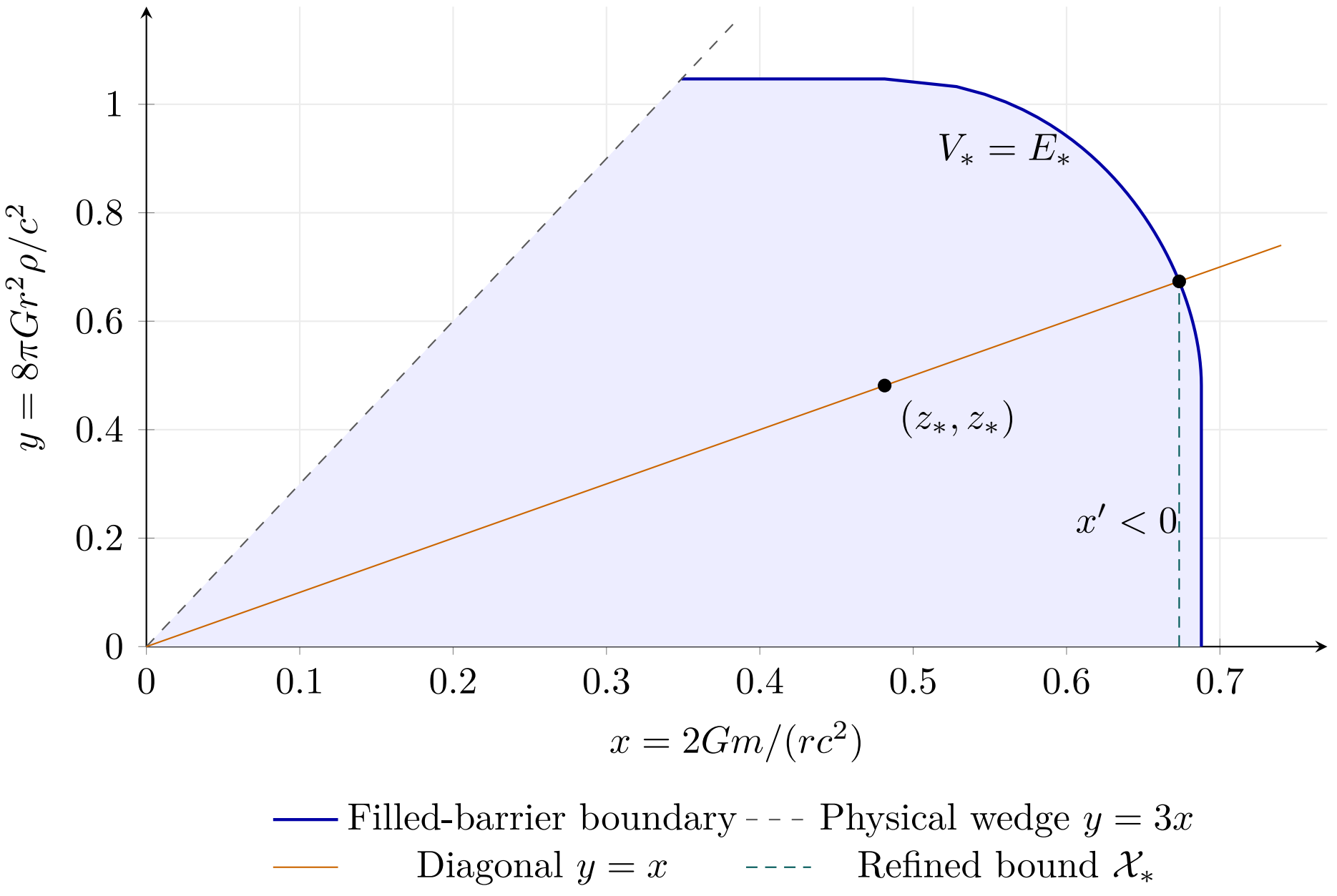
\begin{figure}[htbp]
\centering
\begin{tikzpicture}
\begin{axis}[width=12.5cm,height=7.5cm,xmin=0,xmax=.77,ymin=0,ymax=1.18,
 xlabel={$x=2Gm/(rc^2)$},ylabel={$y=8\pi Gr^2\rho/c^2$},
 axis lines=left,grid=major,grid style={gray!15},
 tick label style={font=\small},label style={font=\small},
 legend style={font=\small,at={(.5,-.22)},anchor=north,draw=none,legend columns=2}]
\addplot[forget plot,draw=none,fill=blue!7] coordinates {(0,0) (.348915260,1.046745781) (.481382468,1.046745781) (0.481382468,1.046745781) (0.528251904,1.032611698) (0.546616265,1.018477615) (0.560247348,1.004343533) (0.571411311,0.990209450) (0.580987593,0.976075367) (0.589427832,0.961941284) (0.597000537,0.947807201) (0.603880696,0.933673119) (0.610189793,0.919539036) (0.616016164,0.905404953) (0.621426344,0.891270870) (0.626471845,0.877136787) (0.631193439,0.863002704) (0.635623967,0.848868622) (0.639790273,0.834734539) (0.643714550,0.820600456) (0.647415318,0.806466373) (0.650908145,0.792332290) (0.654206181,0.778198208) (0.657320568,0.764064125) (0.660260759,0.749930042) (0.663034761,0.735795959) (0.665649323,0.721661876) (0.668110092,0.707527793) (0.670421726,0.693393711) (0.672587984,0.679259628) (0.674611802,0.665125545) (0.676495334,0.650991462) (0.678239995,0.636857379) (0.679846476,0.622723296) (0.681314756,0.608589214) (0.682644091,0.594455131) (0.683832998,0.580321048) (0.684879224,0.566186965) (0.685779696,0.552052882) (0.686530461,0.537918800) (0.687126599,0.523784717) (0.687562125,0.509650634) (0.687829857,0.495516551) (0.687921252,0.481382468) (.687921252,0) (0,0)};
\addplot[blue!65!black,thick] coordinates {(.348915260,1.046745781) (.481382468,1.046745781) (0.481382468,1.046745781) (0.528251904,1.032611698) (0.546616265,1.018477615) (0.560247348,1.004343533) (0.571411311,0.990209450) (0.580987593,0.976075367) (0.589427832,0.961941284) (0.597000537,0.947807201) (0.603880696,0.933673119) (0.610189793,0.919539036) (0.616016164,0.905404953) (0.621426344,0.891270870) (0.626471845,0.877136787) (0.631193439,0.863002704) (0.635623967,0.848868622) (0.639790273,0.834734539) (0.643714550,0.820600456) (0.647415318,0.806466373) (0.650908145,0.792332290) (0.654206181,0.778198208) (0.657320568,0.764064125) (0.660260759,0.749930042) (0.663034761,0.735795959) (0.665649323,0.721661876) (0.668110092,0.707527793) (0.670421726,0.693393711) (0.672587984,0.679259628) (0.674611802,0.665125545) (0.676495334,0.650991462) (0.678239995,0.636857379) (0.679846476,0.622723296) (0.681314756,0.608589214) (0.682644091,0.594455131) (0.683832998,0.580321048) (0.684879224,0.566186965) (0.685779696,0.552052882) (0.686530461,0.537918800) (0.687126599,0.523784717) (0.687562125,0.509650634) (0.687829857,0.495516551) (0.687921252,0.481382468) (.687921252,0)};
\addlegendentry{Filled-barrier boundary}
\addplot[gray!70!black,dashed,domain=0:.383,samples=2] {3*x};
\addlegendentry{Physical wedge $y=3x$}
\addplot[orange!80!black,domain=0:.74,samples=2] {x};
\addlegendentry{Diagonal $y=x$}
\addplot[teal!70!black,densely dashed] coordinates {(.673438593,0) (.673438593,.673438593)};
\addlegendentry{Refined bound $\mathcal X_*$}
\addplot[forget plot,only marks,mark=*,mark size=1.6pt,black] coordinates {(.481382468,.481382468) (.673438593,.673438593)};
\node[anchor=north west,font=\small] at (axis cs:.481382468,.481382468) {$(z_*,z_*)$};
\node[anchor=south,font=\small] at (axis cs:.56,.86) {$V_*=E_*$};
\node[anchor=north east,font=\small] at (axis cs:.682,.29) {$x'<0$};
\end{axis}
\end{tikzpicture}
\caption{The filled barrier in the physical wedge for $k=1/\sqrt3$.
Its upper Lyapunov arc joins the horizontal cap to the vertical side.
The original lower Lyapunov arc is filled in. The upper arc meets
$y=x$ at $(\mathcal X_*,\mathcal X_*)$; to its right, every admissible
point has $y<x$, and therefore $x'<0$. The plotted arc solves its
implicit equation numerically; the proof uses the exact potentials.}
\label{fig:barrier}
\end{figure}

\section{Vacuum behavior and the failure of global Lyapunov decay}

Suppose $R<\infty$, $M=m(R)>0$, and $\rho(R)=0$. From
\eqref{eq:hevolution}, in the physical radius variable,
\begin{equation}
 \frac{\mathrm dH}{\mathrm dr}
 =-\frac{G(m+4\pi r^3P/c^2)}{c^2r^2(1-x)}.
\end{equation}
Hence with
\begin{equation}
 g_R=\frac{GM}{c^2R^2(1-2GM/(Rc^2))}>0
\end{equation}
we have
\begin{equation}
 H(r)\sim g_R(R-r),\qquad
 u(r)\sim\frac{g_R}{\kappa}(R-r).
\end{equation}
The density and pressure obey
\begin{align}
 \rho(r)&\sim\rho_0(g_R/\kappa)^{5/2}(R-r)^{5/2},\\
 P(r)&\sim P_0(g_R/\kappa)^{7/2}(R-r)^{7/2}.
\end{align}
These asymptotics are compatible with a finite surface but do not,
by themselves, prove that all regular solutions reach one.

As $y\to0$, the unmodified $J_k(y)$ diverges to infinity. Therefore a
finite-radius orbit cannot remain in a fixed finite sublevel of
$V_k=U_k+J_k$ all the way to its vacuum endpoint. By contrast,
$J_k(\max\{y,z\})=0$ once $y\leq z$. The filled barrier remains
finite, and its right boundary points inward because $x'<0$ there.
This is the precise distinction between global Lyapunov decay and
the one-sided control needed for the mass--radius bound.

\section{Consequences and limits of the argument}

The result uses neither an autonomous nonlinear phase portrait nor
the assertion that a RAR orbit is ordered beneath a single linear
orbit. It uses pointwise vertical-field domination, a cap provided by
the physical wedge, and the sign of the Lyapunov derivative on its
upper arc. The filled barrier accounts for possible returns after
lower exits, so no assertion about the number of zeros of $y-x$ is
needed.

The constants are independent of $\kappa$, $\chi$, particle mass, and
central density within the stated family. This uniformity is obtained
at the cost of using a broad moment envelope. Sharper estimates of
$C$ in terms of $A,B$ for the cutoff distribution, or a comparator
adapted to density, could improve the compactness constant.

The finite-surface conclusion is a mass--radius restriction for a
configuration already known to exist. It does not establish existence
at a prescribed mass and radius, stability, or the exclusion of other
stronger compactness estimates under additional hypotheses. In
particular, the mere failure of \eqref{eq:failed} does not exclude a
different proof of a stronger bound.

\section*{Acknowledgments}
\addcontentsline{toc}{section}{Acknowledgments}

The kinetic-moment lemma (Lemma~\ref{lem:derivatives}) is due to
ChatGPT (Astra). AI tools also assisted with the LaTeX typesetting
of this manuscript.

\appendix
\section{Correction of the logarithmic coefficient}\label{app:correction}

Slides~30--31 of \cite{ETAMM}, and the explicit expression in
Corollary~3 of the arXiv version \cite{SB-cloud}, use the coefficient
\begin{equation}
 \widetilde b_k=\frac{1+k}{2k}
       \frac{4k+(1+k)^3}{4k+(1+k)^2}
\end{equation}
in the $x$-potential. The system and the general Lyapunov construction
instead require the coefficient $b_k$ in \eqref{eq:HB}. This follows
without numerical calculation from
\begin{equation}
 (z\bk-\ak)(x)=\frac{h_k(x-z)}{1-x},
 \qquad
 \frac{\mathrm d}{\mathrm dx}
       [-h_kx-b_k\log(1-x)]
       =\frac{h_k(x-z)}{1-x}.
\end{equation}
The discrepancy is
\begin{equation}\label{eq:bdiscrepancy}
 \widetilde b_k-b_k
 =\frac{(1+k)(1-k)^2}{2(1+6k+k^2)}.
\end{equation}
It vanishes at $k=1$, but not at $k=1/\sqrt3$ or $k=1/3$.
In particular, the coefficient used here agrees with slide~32:
$b_k=z\delta_k$.

There is a direct diagnostic. If $\widetilde U_k$ is normalized to zero
at $z$ using $\widetilde b_k$, then
\[
 \widetilde U_k'(z)=\frac{\widetilde b_k-b_k}{1-z}\ne0
 \quad(k\ne1).
\]
It therefore cannot be the normalized $x$-potential of the Lyapunov
function centered at $(z,z)$. The quadratic decay identity
\eqref{eq:lyapid} would fail for that substituted potential.

The earlier numerical value $0.6813289023$ at $k=1/\sqrt3$ is the
result of using $\widetilde b_k$ in the inversion. With the consistent
coefficient, the Lambert extent is $0.6879212520$, and the diagonal
refinement is $0.6734385928$. For reference, the same corrected
calculation for a genuinely linear comparator gives
\begin{center}
\begin{tabular}{@{}lrr@{}}
\toprule
$k$ & $X_k^{\mathrm L}$ & $\mathcal X_k$\\
\midrule
$1/3$ & $0.6391177015$ & $0.6235441854$\\
$1/\sqrt3$ & $0.6879212520$ & $0.6734385928$\\
$1$ & $0.6934159640$ & $0.6788359175$\\
\bottomrule
\end{tabular}
\end{center}
The $k=1/3$ row is not transferred to the cutoff EOS by this paper,
because the required global comparison is unavailable. The coefficient
correction concerns the explicit linear formula, not the general
Lyapunov identity or the geometric idea of \cite{SB-cloud}.

\clearpage
\section{The relativistic Michie--King/RAR equation of state from slide 34}
\label{app:slide34}

This appendix reproduces the content of slide~34 of \cite{ETAMM}
(the final overlay is physical PDF page~62), with the pressure
normalization and the asymptotic constants written explicitly.
To avoid confusing the slide's integration variable $x$ and cutoff
variable $y$ with the TOV phase coordinates, they are denoted here by
$\xi$ and $u$, respectively. The slide's particle mass $m$ and Planck
constant $h$ are denoted by $m_f$ and $h_{\mathrm P}$.
Thus the notation changes, but the kinetic formulas do not.

\subsection{Density and pressure integrals}

Fix $\kappa>0$ and $0\leq\chi<\infty$. For $0<u<1/\kappa$, define
the occupation factor on the cutoff interval by
\begin{equation}\label{eq:appoccupation}
 \mathcal F(\xi,u)=\frac{1-\exp(\xi-u)}{1+\chi\exp(\xi-u)},
 \qquad 0\leq\xi\leq u.
\end{equation}
Then the energy--mass density is
\begin{equation}\label{eq:appdensity}
\begin{split}
 \rho(u)={}&
 \frac{8\sqrt2\pi c^3m_f^4\kappa^{3/2}}
      {h_{\mathrm P}^3(1-\kappa u)^{3/2}}
 \int_0^u \xi^{1/2}
 \left(1+\frac{\kappa\xi}{1-\kappa u}\right)^2\\[-2pt]
 &\hspace{15mm}\times
 \left(1+\frac{\kappa\xi}{2(1-\kappa u)}\right)^{1/2}
 \frac{1-\exp(\xi-u)}{1+\chi\exp(\xi-u)}\dd\xi.
\end{split}
\end{equation}
The corresponding pressure is
\begin{equation}\label{eq:apppressure}
 P(u)=\eta\left(\frac{\kappa}{1-\kappa u}\right)^{5/2}
 \int_0^u\xi^{3/2}
 \left(2+\frac{\kappa\xi}{1-\kappa u}\right)^{3/2}
 \frac{1-\exp(\xi-u)}{1+\chi\exp(\xi-u)}\dd\xi,
\end{equation}
where, for spin degeneracy two,
\begin{equation}\label{eq:appeta}
 \eta=\frac{8\pi m_f^4c^5}{3h_{\mathrm P}^3}.
\end{equation}
These are exactly \eqref{eq:sliderho}--\eqref{eq:slidep} with
$\beta=\kappa/(1-\kappa u)$ substituted explicitly.
The parameter $\kappa$ in these kinetic formulas is distinct from the
linear-comparator parameter $k$ used in the main proof.

\subsection{The dilute endpoint and the exponent \texorpdfstring{$7/5$}{7/5}}

For $u\downarrow0$, uniformly on $0\leq\xi\leq u$,
\begin{equation}
 \mathcal F(\xi,u)=\frac{u-\xi}{1+\chi}+O(u^2),
 \qquad \frac{\kappa}{1-\kappa u}\longrightarrow\kappa.
\end{equation}
Consequently the leading density integral is the one displayed on
slide~34:
\begin{equation}\label{eq:appsmallrhointegral}
 \rho(u)\sim
 \frac{8\sqrt2\pi m_f\kappa^{3/2}}{1+\chi}
 \left(\frac{m_fc}{h_{\mathrm P}}\right)^3
 \int_0^u\sqrt\xi\,(u-\xi)\dd\xi
 =\rho_0u^{5/2}.
\end{equation}
Similarly,
\begin{equation}
 P(u)\sim\frac{\eta\,2^{3/2}\kappa^{5/2}}{1+\chi}
 \int_0^u\xi^{3/2}(u-\xi)\dd\xi=P_0u^{7/2}.
\end{equation}
Evaluating the two elementary integrals gives the explicit constants
\begin{equation}\label{eq:appsmallconstants}
 \rho_0=\frac{32\sqrt2\pi m_f^4c^3\kappa^{3/2}}
                  {15h_{\mathrm P}^3(1+\chi)},\qquad
 P_0=\frac{64\sqrt2\pi m_f^4c^5\kappa^{5/2}}
                  {105h_{\mathrm P}^3(1+\chi)}.
\end{equation}
Here $P_0$ is the constant denoted by $p_0$ on the slide.
Since $\rho(u)$ is strictly increasing, elimination of $u$ yields
\begin{equation}\label{eq:apppolytropic}
 \boxed{\displaystyle
 P(\rho)\sim P_0\left(\frac{\rho}{\rho_0}\right)^{7/5}
 \quad(\rho\downarrow0).}
\end{equation}
The constants agree with \eqref{eq:smallrho}--\eqref{eq:smallp}.
In particular, $P_0/(c^2\rho_0)=2\kappa/7$.

\subsection{The relativistic endpoint}

Let $\delta=1-\kappa u\downarrow0$. The coefficient shown on slide~34
is
\begin{equation}\label{eq:apprhoinfty}
 \rho_\infty=
 \frac{8\pi c^3m_f^4}{h_{\mathrm P}^3}
 \int_0^1 v^3
 \frac{1-\exp((v-1)/\kappa)}{1+\chi\exp((v-1)/\kappa)}\dd v>0.
\end{equation}
The integration variable $v$ replaces the slide's $w$ to distinguish
it from the pressure-to-density ratio of the main text. Then
\begin{equation}\label{eq:apphighasymptotics}
 \rho(u)\sim\frac{\rho_\infty}{(1-\kappa u)^4},\qquad
 P(u)\sim\frac{c^2\rho_\infty}{3(1-\kappa u)^4}
 \quad(u\uparrow1/\kappa).
\end{equation}

To verify the common coefficient, set $q=\beta\xi$ and then
$v=\delta q$ in the energy representation
\eqref{eq:rhoenergy}--\eqref{eq:penergy}. The interval becomes
$0\leq v\leq1-\delta$, the occupation factor becomes
$F((v-1+\delta)/\kappa)$, and
\begin{align}
 \delta^4\rho(u)
 &=\kin\int_0^{1-\delta}
 F((v-1+\delta)/\kappa)(v+\delta)^2
                 \sqrt{v(v+2\delta)}\dd v,\\
 \delta^4P(u)
 &=\frac{\kin c^2}{3}\int_0^{1-\delta}
 F((v-1+\delta)/\kappa)[v(v+2\delta)]^{3/2}\dd v.
\end{align}
Extend the integrands by zero to $[0,1]$. They are uniformly bounded
for sufficiently small $\delta$ and converge pointwise almost
everywhere to $v^3F((v-1)/\kappa)$. Dominated convergence proves
\eqref{eq:apphighasymptotics} and, on eliminating $u$, gives
\begin{equation}\label{eq:apphighEOS}
 \boxed{\displaystyle P(\rho)\sim\frac{c^2}{3}\rho
 \quad(\rho\to\infty).}
\end{equation}
The two endpoint laws in this appendix are those of slide~34.
For the global derivative inequality see Proposition~\ref{prop:kinetic};
Appendix~\ref{app:hydrodynamic} gives a direct density--pressure proof.

\clearpage
\section{Direct density--pressure estimates in enthalpy variables}
\label{app:hydrodynamic}

This appendix gives a direct proof of the slope bound \eqref{eq:bounds}
from the two density and pressure integrals. The statement is in
the hydrodynamic variables \(\rho,\Pi,H\), where \(\Pi=P/c^2\) and
\(H\) is the enthalpy normalized at vacuum. The proof keeps the
original integral kernels and does not introduce separate kinetic moments.

\subsection{The starting formulas of Bors and Sta\'nczy}\label{rar:starting}
We use the density and pressure formulas of Bors and Sta\'nczy~\cite[\S5]{BS-TOV}, with the normalization of Appendix~\ref{app:slide34}. The density \(\rho\) is the energy density divided by \(c^2\); \(P\) is the physical pressure. Thus the two variables of the same dimension are \(\rho\) and \(\Pi=P/c^2\).

Let \(m_f>0\) be the particle mass, \(h_{\mathrm P}>0\) Planck's constant, \(k_{\mathrm B}\) Boltzmann's constant, \(T_{\mathrm{phys}}>0\) the local temperature, and \(p=|\mathbf p|\) the momentum magnitude. Write
\[
 \varepsilon(p)=\sqrt{m_f^2c^4+c^2p^2}-m_fc^2,\qquad
 q=\frac{\varepsilon(p)}{m_fc^2}.
\]
With cutoff energy \(\varepsilon_c\), chemical potential \(\mu\), and spin degeneracy two, the occupation and the isotropic integrals are
\begin{align}
 f_c(\mathbf p)
   &=\left(\frac{1-\exp((\varepsilon-\varepsilon_c)/(k_{\mathrm B}T_{\mathrm{phys}}))}
                   {1+\exp((\varepsilon-\mu)/(k_{\mathrm B}T_{\mathrm{phys}}))}
     \right)_+,\label{rar:phasef}\\
 \rho&=\frac{2m_f}{h_{\mathrm P}^3}
     \int_{\mathbb R^3} f_c(\mathbf p)(1+q)\dd^3\mathbf p,\label{rar:phaseR}\\
 P&=\frac{4}{3h_{\mathrm P}^3}
     \int_{\mathbb R^3} f_c(\mathbf p)
       \frac{\varepsilon(1+q/2)}{1+q}\dd^3\mathbf p.\label{rar:phaseP}
\end{align}
The subscript \(+\) denotes the positive part. To keep \(x,y\) available for the TOV phase plane, we write \(\xi,u\) for the variables \(x,y\) in the equation-of-state formulas of \cite[\S5]{BS-TOV}:
\begin{equation}\label{rar:parameter}
 \xi=\frac{\varepsilon}{k_{\mathrm B}T_{\mathrm{phys}}},\qquad
 u=\frac{\varepsilon_c}{k_{\mathrm B}T_{\mathrm{phys}}},\qquad
 \beta=\frac{k_{\mathrm B}T_{\mathrm{phys}}}{m_fc^2}
      =\frac{\kappa}{1-\kappa u},\qquad
 \chi=\exp\left(\frac{\varepsilon_c-\mu}{k_{\mathrm B}T_{\mathrm{phys}}}\right).
\end{equation}
The temperature relation is \(1-\kappa u=T_R/T_{\mathrm{phys}}\), where \(\kappa=k_{\mathrm B}T_R/(m_fc^2)>0\) and \(T_R\) is the reference temperature at the cutoff surface in that parametrization. We fix \(\kappa\) and \(\chi\); the formulas also allow the limit \(\chi=0\). Throughout the interior of the family,
\[
  0<u<1/\kappa,\qquad
  \mathcal F(\xi,u)=\frac{1-e^{\xi-u}}{1+\chi e^{\xi-u}}\quad(0\le \xi\le u).
\]
In particular \(\mathcal F(u,u)=0\) and \(\mathcal F(\xi,u)>0\) for \(0<\xi<u\).

The change of variables is \(q=\beta\xi\). Define the two factors
\[
 L=1+\beta\xi,\qquad S=\sqrt{\beta\xi(2+\beta\xi)}=\sqrt{L^2-1}.
\]
Then \(p=m_fcS\) and \(p^2\,dp=m_f^3c^3\beta LS\,d\xi\). To see the pressure normalization, use \(\varepsilon(1+q/2)=m_fc^2 S^2/2\); the factor \(L^{-1}\) in \eqref{rar:phaseP} cancels the \(L\) in the measure \(4\pi p^2\,dp\). With
\[
 \kin{}=\frac{8\pi m_f^4c^3}{h_{\mathrm P}^3},\qquad
 \eta=\frac{\kin{}c^2}{3}=\frac{8\pi m_f^4c^5}{3h_{\mathrm P}^3},
\]
one obtains the density and pressure integrals of \cite[\S5]{BS-TOV}:
\begin{align}
 \rho(u)&=\sqrt{2}\,\kin{}\beta^{3/2}
    \int_0^u \xi^{1/2}(1+\beta\xi)^2
                 (1+\beta\xi/2)^{1/2}\mathcal F(\xi,u)\dd\xi,\label{rar:publishedR}\\
 P(u)&=\eta\,\beta^{5/2}
    \int_0^u \xi^{3/2}(2+\beta\xi)^{3/2}
                 \mathcal F(\xi,u)\dd\xi.\label{rar:publishedP}
\end{align}
In particular the two formulas use the same normalization after writing \(\Pi=P/c^2\). For the proof it is useful to group the existing integrands as
\begin{equation}\label{rar:kernels}
 \rho=\int_0^u \mathcal F\mathcal R\dd\xi,\qquad
 \Pi=\int_0^u \mathcal F\mathcal Q\dd\xi,\qquad
 \mathcal R=\kin{}\beta L^2S,\quad \mathcal Q=\frac{\kin{}\beta}{3}S^3 .
\end{equation}
Here \(\mathcal R,\mathcal Q\) are exactly the density and normalized pressure kernels with their common occupation factor removed.

\subsection{Density--pressure lemma}\label{rar:lemma-sec}
Use the parameter
\begin{equation}\label{rar:H}
 H=-\log(1-\kappa u),\qquad
 u=\frac{1-e^{-H}}{\kappa},\qquad
 \beta=\kappa e^H,\qquad H_u=\beta.
\end{equation}
We first use \(H\) as a change of variable; the lemma identifies it with the enthalpy normalized to zero at vacuum.

\begin{lemma}[Fluid form of the kinetic estimate]\label{rar:lemma}
Fix \(\kappa>0\), \(0\le\chi<\infty\), and positive physical constants. Let \(\rho\) and \(\Pi\) be given by \eqref{rar:publishedR}--\eqref{rar:publishedP}. For every \(H>0\) the functions are smooth and satisfy
\begin{equation}\label{rar:main}
 \boxed{\quad
   0<3\Pi<\rho,\qquad \Pi_H=\rho+\Pi,\qquad
   \rho_H>3\rho+\frac{\rho^2}{3\Pi}.
 \quad}
\end{equation}
Equivalently the density inequality is \(3\Pi(\rho_H-3\rho)>\rho^2\).
In the original cutoff parameter it reads
\begin{equation}\label{rar:cutoffbounds}
 P_u=\beta(c^2\rho+P),\qquad
 \rho_u>\beta\left(3\rho+\frac{c^2\rho^2}{3P}\right).
\end{equation}
\end{lemma}

\begin{proof}
For \(0<\xi<u\) the ratio of the kernels is
\begin{equation}\label{rar:kernelratio}
 \frac{3\mathcal Q}{\mathcal R}=\frac{S^2}{L^2}=1-L^{-2}\in(0,1).
\end{equation}
Both kernels and the occupation are positive there, so \(0<3\Pi<\rho\).

\emph{Differentiation at a moving cutoff.}
For \(u\) in a compact interval inside \((0,1/\kappa)\), \(\beta\) and its derivatives are bounded and \(\beta\) is bounded away from zero. The formula for \(\mathcal F\) extends smoothly to any finite rectangle in \((\xi,u)\); its extension is used only to apply Leibniz's rule. Since
\[
 S=\xi^{1/2}\sqrt{\beta(2+\beta\xi)},
\]
each kernel is a smooth factor times \(\xi^{1/2}\) or \(\xi^{3/2}\), including at \(\xi=0\) after removal of that power. The same fixed powers control their \(u\) derivatives locally uniformly. Thus the integrals may be differentiated in \(u\), and they are smooth in the interior: one may also see smoothness of every order by writing \(\xi=ut\), so the only endpoint powers are those of \(t\) on a fixed unit interval.

For \(D=\mathcal R\) or \(\mathcal Q\), use \(\mathcal F_u=-\mathcal F_\xi\) and \(\beta_u=\beta^2\). In fact \(D\) has the form \(\kin{}\beta\,d(1+\beta\xi)\) for a function \(d\) of one argument. Differentiation for \(\xi>0\) gives
\begin{equation}\label{rar:transport}
 D_u=\beta(D+\xi D_\xi)=\beta\,\partial_\xi(\xi D).
\end{equation}
There is no contribution from the moving upper limit because \(\mathcal F(u,u)=0\). Integrating the term \(-\mathcal F_\xi D\) by parts on \([\epsilon,u]\), then taking \(\epsilon\downarrow0\), gives
\begin{align}
 \frac{d}{du}\int_0^u \mathcal F D\dd\xi
  &=\int_0^u (\mathcal F_u D+\mathcal F D_u)\dd\xi \notag\\
  &=\int_0^u \mathcal F(D_\xi+D_u)\dd\xi
   =\int_0^u \mathcal F\,\partial_\xi(LD)\dd\xi .\label{rar:integralderiv}
\end{align}
Here the lower boundary term tends to zero since \(\mathcal F\) is bounded and \(D(\xi,u)\to0\). The derivative of \(LD\) is \(O(\xi^{-1/2})\) for density and \(O(\xi^{1/2})\) for pressure, so the limiting integrals converge.

\emph{Exact derivative formulas.}
From \(L_\xi=\beta\) and \(S_\xi=\beta L/S\), the two weighted kernels satisfy
\begin{align}
 \partial_\xi(L\mathcal Q)
  &=\frac{\kin{}\beta^2}{3}\big(S^3+3L^2S\big)
    =\beta(\mathcal R+\mathcal Q),\label{rar:pressurekerder}\\
 \partial_\xi(L\mathcal R)
  &=\kin{}\beta^2\left(3L^2S+\frac{L^4}{S}\right)
    =\beta\left(3\mathcal R+\frac{\mathcal R^2}{3\mathcal Q}\right).\label{rar:densitykerder}
\end{align}
Divide \eqref{rar:integralderiv} by \(H_u=\beta\). This proves
\begin{equation}\label{rar:exact}
 \Pi_H=\rho+\Pi,\qquad
 \rho_H-3\rho=\int_0^u \mathcal F\,\frac{\mathcal R^2}{3\mathcal Q}\dd\xi .
\end{equation}
The second identity expresses a derivative of the original density. It does not require an additional quantity to specify the equation of state.

\emph{Strict inequality.}
Cauchy--Schwarz in the measure \(\mathcal F(\xi,u)\,d\xi\) gives
\[
 \left(\int_0^u \mathcal F\mathcal R\dd\xi\right)^2
 \le
 \left(\int_0^u \mathcal F\,3\mathcal Q\dd\xi\right)
 \left(\int_0^u \mathcal F\,\frac{\mathcal R^2}{3\mathcal Q}\dd\xi\right).
\]
Equivalently, the exact excess above the lower bound is a square:
\begin{equation}\label{rar:gap}
 \boxed{\displaystyle
 \rho_H-3\rho-\frac{\rho^2}{3\Pi}
  =\int_0^u\frac{\mathcal F}{3\mathcal Q}
       \left(\mathcal R-\frac{\rho}{\Pi}\mathcal Q\right)^{\!2}\dd\xi .}
\end{equation}
Expanding the right side uses only \eqref{rar:kernels} and \eqref{rar:exact}. A zero right side would force the ratio \(\mathcal R/\mathcal Q\) to be constant almost everywhere on the occupied interval. But \(1-L^{-2}\) has derivative \(2\beta/L^3>0\) and \(\mathcal F>0\) on \((0,u)\). This proves strictness.
\end{proof}

\subsection{Enthalpy, compressibility, and the slope bound}
The integrals tend to zero as \(u\downarrow0\) (the leading powers are proved below). Since \(\Pi_H=\rho+\Pi\), the integral from vacuum obeys
\begin{equation}\label{rar:enthalpy}
 H=\int_0^P\frac{\dd p}{c^2\rho(p)+p}
   =\int_0^\Pi\frac{\dd\pi}{\rho(\pi)+\pi}.
\end{equation}
Here \(\rho(p)\) or \(\rho(\pi)\) denotes density along the equation of state at that pressure. The equality follows first with a positive lower endpoint by the substitution \eqref{rar:exact}, and then by taking the lower endpoint to zero; it also proves convergence at vacuum.

Since \(\rho_H>0\), pressure is a single-valued smooth function of density in the positive range. Introduce the dimensionless quantities
\[
 w=\frac{\Pi}{\rho},\qquad a=\frac{\dd\Pi}{\dd\rho}
     =\frac{1}{c^2}\frac{\dd P}{\dd\rho}.
\]
The derivative \(a\) is taken along the fixed-\(\kappa,\chi\) equation of state. The lemma and the chain rule give
\begin{equation}\label{rar:slope}
 0<a=\frac{\rho+\Pi}{\rho_H}
 <\frac{3\Pi(\rho+\Pi)}{\rho(\rho+9\Pi)}
 =\frac{3w(1+w)}{1+9w}<\frac13,\qquad 0<w<\frac13.
\end{equation}
The last inequality is equivalent to \(9w^2<1\). In physical units this is
\[
 0<\frac{\dd P}{\dd\rho}
    <\frac{3P}{\rho}\frac{c^2\rho+P}{c^2\rho+9P}.
\]
Two useful restatements, obtained by inversion and logarithmic differentiation, are
\begin{equation}\label{rar:hydrovariants}
 0<\frac{\dd H}{\dd\rho}<\frac{3\Pi}{\rho(\rho+9\Pi)},
 \qquad
 0<\frac{\dd\log P}{\dd\log\rho}=\frac{a}{w}
   <\frac{3(1+w)}{1+9w}.
\end{equation}
Also \(w_H=1+w-w\,\rho_H/\rho<2(1/3-w)\).
The strict estimate encodes the variation of the ratio \eqref{rar:kernelratio} over the occupied interval. It gives a bound on the local equation-of-state slope along this family.

\subsection{The two endpoint regimes from the same integrals}
\paragraph{Dilute limit.}
Use \(\xi=ut\), \(0\le t\le1\), and let \(u\downarrow0\). Uniformly in \(t\),
\[
 \mathcal F(ut,u)=\frac{u(1-t)}{1+\chi}[1+O(u)],\qquad
 L=1+O(u),\qquad
 S=\sqrt{2\kappa u t}\,[1+O(u)].
\]
The relative expression for \(\mathcal F\) follows by factoring \(u(1-t)\) out of \(1-e^{-u(1-t)}\); it extends to \(t=1\) before multiplication by this factor. Inserting these expressions into \eqref{rar:kernels} and using
\(\int_0^1 t^{1/2}(1-t)\,dt=4/15\) and \(\int_0^1 t^{3/2}(1-t)\,dt=4/35\), one finds
\begin{align}
 \rho(u)&\sim\rho_0u^{5/2},
 &\rho_0&=\frac{4\sqrt2 \kin{}\kappa^{3/2}}{15(1+\chi)},\label{rar:lowR}\\
 P(u)&\sim P_0u^{7/2},
 &P_0&=\frac{8\sqrt2 \kin{}c^2\kappa^{5/2}}{105(1+\chi)}.\label{rar:lowP}
\end{align}
These agree with \eqref{eq:smallrho}--\eqref{eq:smallp} and \eqref{eq:appsmallconstants}. Eliminating \(u\) yields \(P(\rho)\sim P_0(\rho/\rho_0)^{7/5}\). In enthalpy variables \(H\sim\kappa u\), so \(\rho(H)\sim\rho_0(H/\kappa)^{5/2}\) and \(P(H)\sim P_0(H/\kappa)^{7/2}\).

For the slope, apply the same substitution directly to \eqref{rar:exact}:
\[
 \rho_H-3\rho=\kin{}\beta\int_0^u \mathcal F\,\frac{L^4}{S}\dd\xi
 \sim \frac{\kin{}\sqrt\kappa}{\sqrt2(1+\chi)}\,u^{3/2}
             \int_0^1 t^{-1/2}(1-t)\,dt
 = \frac{5\rho_0}{2\kappa}\,u^{3/2}.
\]
The weight \(t^{-1/2}(1-t)\) is integrable and the remaining factor tends to one uniformly away from the harmless endpoints, with a uniform bound on \(0<t<1\). The integral is \(4/3\). Since \(3\rho=O(u^{5/2})\), the same leading term holds for \(\rho_H\). Thus
\begin{equation}\label{rar:lowSlope}
 w\sim \frac{2\kappa}{7}u,\qquad
 a\sim \frac{2\kappa}{5}u,\qquad
 \frac{\dd\log P}{\dd\log\rho}\longrightarrow\frac75 .
\end{equation}
This derivative asymptotic follows from the exact formulas, without differentiating an asymptotic equivalence.

\paragraph{Relativistic limit.}
Set \(\delta=1-\kappa u=e^{-H}\), \(v=\kappa\xi\),
\[
 W_\delta(v)=\frac{1-e^{(v-1+\delta)/\kappa}}
                   {1+\chi e^{(v-1+\delta)/\kappa}},\qquad
 S_\delta(v)=\sqrt{v(v+2\delta)},\qquad 0\le v\le1-\delta .
\]
On this interval \(L=(v+\delta)/\delta\), \(S=S_\delta/\delta\), and \(\beta\,d\xi=\delta^{-1}\,dv\). The exact rescalings are
\begin{align}
 \delta^4\rho &=\kin{}\int_0^{1-\delta} W_\delta(v)(v+\delta)^2 S_\delta(v)\dd v,\label{rar:scaleR}\\
 3\delta^4\Pi &=\kin{}\int_0^{1-\delta} W_\delta(v) S_\delta(v)^3\dd v,\label{rar:scaleP}\\
 \delta^2(\rho-3\Pi)&=\kin{}\int_0^{1-\delta} W_\delta(v) S_\delta(v)\dd v.\label{rar:scaleD}
\end{align}
The difference formula uses \((v+\delta)^2-S_\delta(v)^2=\delta^2\). Extend the integrands by zero to \([0,1]\). Since \(0\le W_\delta\le1\) and the remaining factors are uniformly bounded, dominated convergence gives
\begin{equation}\label{rar:highLaws}
 \rho\sim\rho_\infty\,\delta^{-4},\qquad
 P\sim\frac{c^2\rho_\infty}{3}\,\delta^{-4},\qquad
 \rho_\infty=\kin{}\int_0^1 W_0(v)v^3\dd v>0.
\end{equation}
In particular \(P(\rho)\sim c^2\rho/3\) as \(\rho\to\infty\). The density is strictly increasing and ranges from zero to infinity, so the equation of state is defined on the whole positive density axis.

For a more precise comparison of \(w\) and \(a\), use the relative fluid difference
\begin{equation}\label{rar:relative}
 d=\frac{\rho-3\Pi}{\rho}=1-3w>0,\qquad
 d\sim\frac{\kin{}\int_0^1 W_0(v)v\,dv}{\rho_\infty}\,\delta^2.
\end{equation}
The derivative identity \eqref{rar:exact}, together with \(L^2=S^2+1\), can be rearranged as
\begin{equation}\label{rar:high_deriv}
 \rho_H=4\rho+(\rho-3\Pi)+\kin{}\beta\int_0^u \frac{\mathcal F}{S}\dd\xi .
\end{equation}
The last integral, including its prefactor, is bounded by
\(\kin{}\,\operatorname{arcosh}(1+\beta u)=\kin{}\,\operatorname{arcosh}(\delta^{-1})\),
because \(\mathcal F\le1\) and \(\partial_\xi\operatorname{arcosh}(1+\beta\xi)=\beta/S\).
It is \(O(\log(\delta^{-1}))=o(\rho-3\Pi)\). Consequently
\begin{equation}\label{rar:highSlope}
 \frac{\rho_H}{\rho}=4+d+o(d),\qquad
 w=\frac13-\frac d3,\qquad
 a=\frac{4-d}{3(4+d+o(d))}=\frac13-\frac d6+o(d).
\end{equation}
Both \(w\) and \(a\) approach the radiation value \(1/3\), but their first corrections differ.

\subsection{The TOV equations in enthalpy form}\label{rar:TOV}
Consider a regular static spherical solution on its connected positive-density
branch with \(x<1\). In the notation \eqref{eq:milne}, division of the pressure
equation \eqref{eq:tov} by \(c^2(\rho+\Pi)\) gives
\begin{equation}\label{rar:Hr}
 H_r=-\frac{G(m+4\pi r^3\Pi)}{c^2r^2(1-x)},\qquad u_r=H_r/\beta.
\end{equation}
In particular, with primes denoting \(d/ds=r\,d/dr\), we recover
\(H'=-(x+wy)/(2(1-x))\) and \(x'=y-x\).
Differentiating \(y=8\pi Gr^2\rho/c^2\) and using
\(\rho_H/\rho=(1+w)/a\) gives the direct enthalpy form of
\eqref{eq:nonlinear}:
\begin{equation}\label{rar:phaseY}
 y'=y\left[2-\frac{\rho_H}{2\rho}\frac{x+wy}{1-x}\right]
   =y\left[2-\frac{1+w}{2a}\frac{x+wy}{1-x}\right]
   =F_{\mathrm R}(x,y;\rho).
\end{equation}
By \eqref{rar:Hr}, \(H_r<0\) at positive radius in this branch. Since
\(\rho_H>0\), the density is strictly decreasing. Thus
\(m(r)=4\pi\int_0^r t^2\rho(t)\,dt>4\pi r^3\rho(r)/3\), which gives
the physical wedge \(0<y<3x\). Continuity of \(\rho\) at a regular center
gives \(m(r)/r^3\to4\pi\rho_c/3\), in agreement with the center
asymptotics \eqref{eq:center}.

If the branch reaches a finite vacuum surface \(R\) with \(M=m(R)>0\)
and \(x(R)<1\), equation \eqref{rar:Hr} gives
\[
 -H_r\longrightarrow
 g_R=\frac{GM}{c^2R^2(1-2GM/(Rc^2))}>0,\qquad
 H(r)\sim g_R(R-r).
\]
Using \(u\sim H/\kappa\) in \eqref{rar:lowR}--\eqref{rar:lowP} yields
\begin{equation}\label{rar:boundary}
 \rho(r)\sim\rho_0(g_R/\kappa)^{5/2}(R-r)^{5/2},\qquad
 P(r)\sim P_0(g_R/\kappa)^{7/2}(R-r)^{7/2}.
\end{equation}
These are the conditional surface laws of the main text.

\subsection{A uniform vertical comparison}
For the linear equation of state \(\Pi=k\rho\), \(k>0\), set \(w=a=k\) in \eqref{rar:phaseY}:
\[
 F_k(x,y)=y\left[2-\frac{1+k}{2k}\frac{x+ky}{1-x}\right].
\]
At \(0<x<1\), \(y>0\), set \(t=y/x\). Cancelling positive factors gives the exact test
\begin{equation}\label{rar:comparisonTest}
 F_{\mathrm R}\le F_k
 \quad\Longleftrightarrow\quad
 a\le\frac{k(1+w)}{1+k}\frac{1+wt}{1+kt}.
\end{equation}
If \(w\le k\), the derivative of the last fraction in \(t\) is \((w-k)/(1+kt)^2\le0\). Thus \(t=3\) gives the strongest requirement in the wedge \(0<t\le3\). By \eqref{rar:slope} it suffices that
\begin{equation}\label{rar:gtest}
 \frac{3w}{(1+9w)(1+3w)}\le \frac{k}{(1+k)(1+3k)}.
\end{equation}
The two sides are \(g(3w)\) and \(g(k)\) for \(g(v)=v/[(1+v)(1+3v)]\). Its derivative has the sign of \(1-3v^2\), so its unique maximum on \(v>0\) occurs at \(1/\sqrt3\). That maximizer also belongs to the interval \(0<3w<1\). Hence the value
\[
 \boxed{\ks=1/\sqrt3}
\]
is optimal for this particular sufficient test over the entire pressure-ratio interval. Since \(w<1/3<\ks\) and \eqref{rar:slope} is strict, \(F_{\mathrm R}<F_{\ks}\) at every positive-density state in the physical wedge. This is a field comparison and does not identify the RAR trajectory with a linear trajectory.

\paragraph{Why the radiation value does not give this comparison.}
For \(k=1/3\), the test at \(t=3\) requires \(a\le(1+w)(1+3w)/8\). From \eqref{rar:relative}--\eqref{rar:highSlope},
\[
 \frac{(1+w)(1+3w)}8=\frac13-\frac d4+O(d^2),
 \qquad
 a-\frac{(1+w)(1+3w)}8=\frac d{12}+o(d)>0
\]
at sufficiently high density. Thus equality of the limiting ratios does not imply a uniform radiation comparison over the wedge. Since the inequality fails at \(t=3\) strictly at these states, it also fails sufficiently close to \(t=3\) from below at the same states.

\subsection{Consequence for the filled barrier}
The enthalpy proof recovers both \eqref{eq:bounds} and the strict
comparison \eqref{eq:domination}. By \eqref{rar:phaseY} and the regular-center
limits, the hypotheses of Lemma~\ref{lem:barrier} are therefore met
on the initially horizon-free positive-density branch. In the
notation of Section~\ref{sec:barrier}, put \(k=\ks\). The orbit starts
at \(\Ps_{\ks}=0\); the inward directions on the horizontal cap,
upper arc, and right side keep it below \(E_*\). The same phase region
is displayed in Figure~\ref{fig:barrier}.

To see why the diagonal bound is the one relevant to compactness,
suppose \(x\) first reaches \(\Xs\) from below. Then \(x'\geq0\)
forces \(y\geq x=\Xs>z_*\). Monotonicity of \(J_*\) there and the defining
equation \eqref{eq:mainroot} would imply \(\Ps_{\ks}\geq E_*\),
contradicting the strict sublevel. Thus Proposition~\ref{prop:diagonal}
gives precisely \eqref{eq:mainbound}; no separate kinetic input is needed
for this application once the density--pressure lemma is established.
If the branch reaches a finite vacuum surface, \(y\to0\) and \(x(R)>0\)
give \(x'<0\) just inside the surface. The strict bound then persists
at \(R\), yielding the mass--radius inequality \eqref{eq:radius}.

\end{document}